\documentclass[aps,prb,groupedaddress,twocolumn,showpacs,10pt,floatfix]{revtex4-1}

\usepackage{amsmath}
\usepackage{graphicx}
\usepackage{dcolumn}
\usepackage{bm}
\begin{document}


\title{Influence of nonmagnetic dielectric spacers on the spin wave response \\ of  one-dimensional planar magnonic crystals}


\author{G.~Centa{\l}a$^1$}
\author{M.\,L.~Sokolovskyy$^2$}
\author{C. S. Davies$^{3,4}$}
\author{M.~Mruczkiewicz$^5$}
\author{S.~Mamica$^1$}
\author{J.~Rych{\l}y$^{1,7}$}
\author{J.~W.~K{\l}os$^1$}
\email{klos@amu.edu.pl}
\author{V. V.~Kruglyak$^6$}
\author{M.~Krawczyk$^1$}

\affiliation{$^1$Faculty of Physics, Adam Mickiewicz University in Poznan, Uniwersytetu Poznanskiego 2, Pozna\'{n}, 61-614, Poland}

\affiliation{$^2$43a Eilat str., Holon, 5843045, Israel}

\affiliation{$^3$FELIX Laboratory, Radboud University, 7c Toernooiveld, 6525 ED Nijmegen, The Netherlands}

\affiliation{$^4$Radboud University, Institute for Molecules and Materials, 135 Heyendaalseweg, 6525 AJ Nijmegen, The Netherlands}

\affiliation{$^5$Institute of Electrical Engineering, Slovak Academy of Sciences, Dubravska cesta 9, 841 04 Bratislava, Slovakia}

\affiliation{$^6$University of Exeter, Stocker Road, Exeter, EX4 4QL, U.K.}

\affiliation{$^7$Institute of Molecular Physics Polish Academy of Sciences, Mariana Smoluchowskiego 17, 60-179 Poznań, Poland}

\date{\today}

\begin{abstract}
One-dimensional planar magnonic crystals are usually fabricated as a sequence of stripes intentionally or accidentally separated by non-magnetic spacers. The influence of spacers on shaping the spin wave spectra is complex and still not completely clarified. We perform detailed numerical studies of the one-dimensional single- and bi-component magnonic crystals comprised of a periodic array of thin  ferromagnetic stripes separated by non-magnetic spacers.  We show that the dynamic dipolar interactions between the stripes, mediated even by ultra-narrow non-magnetic spacers, lead to a significant increase in the frequency of the ferromagnetic resonance mode, while simultaneously reducing the spin wave group velocity. We attribute these spectral deformations to the modifications of dipolar pinning and shape anisotropy, both of which are dependent on the width of the spacers and the thickness of the stripes,  as well as changes with the dynamical magnetic volume charges formed due to inhomogeneous spin wave amplitude.

\end{abstract}

\pacs{75.75.+a,76.50.+g,75.30.Ds,75.50.Bb}

 \maketitle


\section{Introduction\label{intro}}

Periodic magnetic  structures used as a medium for the controlled propagation of spin waves (SWs) are called magnonic crystals (MCs).\cite{Vasseur96,Puszkarski03} The spectrum of an MC is similar to any other structure featuring discrete translational symmetry (e.g. photonic crystals\cite{Joannopoulos08} or phononic crystals\cite{Khelif16}), and is strongly influenced by the presence of magnonic band gaps, in which there are no allowed magnonic states.\cite{Krawczyk08} One of the first attempts to study the propagation of SWs in periodic magnetic structures was made by Elachi.\cite{Elachi76} Recently, the number of studies on this topic has surged and continues to grow at a fast pace.\cite{Kruglyak10b,Demokritov13,Krawczyk14}

The fabrication process of artificial periodic structures with characteristic dimensions on the nanoscale is very hard to reliably control. It is especially difficult to control the quality of the lateral surfaces and interfaces between adjacent materials that constitute the MC. The roughness of lateral surfaces in planar nanostructures is usually larger than that of the top surfaces and bottom interfaces \cite{Zhao14} mostly due to the differences in their formation during fabrication. The horizontal surfaces or interfaces of nanodots and nanostripes are usually formed during the deposition of continuous layers of the material, whereas the lateral (inter)faces are formed by patterning techniques. \cite{Adeyeye08} The roughness can affect the surface anisotropy due to geometrical factors \cite{Bruno88} or because of the changes in the chemisorption (oxidation). \cite{Anderson71} The impact of the surface anisotropy on the 
SW pinning can be different on lateral and on horizontal faces due to large difference in their areas for planar structures. Moreover, the diffusion may occur  between homogeneous materials, giving rise to a transition layer with properties different from the original constituents. This diffusion is proportional to the concentration gradient at the interface, and so, it is most rapid in structures prepared with sharp interfaces. 

At the same time, most theoretical papers investigating the SWs in MCs consider structures with sharp interfaces.\cite{Vasseur96,Puszkarski03,Kruglyak04,Krawczyk08,Klos12b,Rychly2017,Milinska14} However, in some cases, this assumption is a severe idealization, and other models should be used instead. For example, in Ref.~[\onlinecite{Ignatchenko00}], the coordinate dependence of magnetic parameters of a transition layer was approximated by the Jacobian elliptic sine function, and a strong dependence of the magnonic spectrum and coefficients of reflection and transmission of SWs upon the width of interfaces was demonstrated. The spectrum of exchange-mediated SWs in an MC with diffuse interfaces was also derived for a model with cosine-like\cite{Tkachenko06} and linear\cite{Tkachenko10} profiles of the uniaxial anisotropy at the interfaces. It was suggested that the performance of magnonic devices employing MCs as a filtering element may degrade as the thickness of the interfaces increases due to the interdiffusion between constituent layers of MCs.

In the experimental realization of bi-component MCs \cite{Wang10,Barman2016,Silvani18}, the interfaces between layers can suffer from oxidation, and so, a significant change of magnetic properties near the edges can be expected. This is especially important in the case of ferromagnetic metals because of their high reactivity with oxygen, \cite{Zhu2010} giving rise to oxides displaying paramagnetic or antiferromagnetic properties at temperatures above and below the Neel temperature, respectively. This can lead to the formation of interfacial `spacer' regions that are non-magnetic altogether.  The oxidation on lateral surfaces can be different (usually enhanced) compared to the horizontal surfaces, mostly due to the larger roughness and the difficulties associated with preventing exposure to oxygen (especially during the lithography process). Therefore, it is reasonable to consider the planar MC with the surface anisotropy on the edges of nanopatterned elements (e.g., stripes, dots) but to neglect the impact of the surface anisotropy on the top surfaces.

The purpose of our study is to investigate the influence of such narrow non-magnetic spacers between the edges of two ferromagnetic stripes in `one-dimensional' magonic crystals (1D MCs) on their magnonic band spectrum. The investigated 1D MCs are formed by thin permalloy (and cobalt) stripes, arranged in-plane as shown in Fig.~\ref{Fig:structure}. The term one-dimensional magnonic crystal stresses that the considered planar structure is periodic in one of two  in-plane dimensions only\cite{Chumak08,Baumgaertl18,Gubbiotti19}. Such a study is crucial to understand the physical mechanism responsible for the dynamical coupling of SWs in MCs, and to explain the influence of pinning at the stripe edges and dipolar couplings between constituent materials on the magnonic band structure. 

The geometrical factors and properties at surfaces and edges of the ferromagnetic structure result in different SW pinning conditions for the magnetization and SWs.\cite{Wigen62,Puszkarski74,Guslienko05,Wang2018} In the exchange dominated regime, it is usually justified to consider the impact of the surfaces as a local effect which can be accounted for by the introduction of a phenomenological  parameter. The SW pinning can originate from the surface anisotropy with energy density $K_s$. The parameter $K_s$  describes the additional torque which acts on the magnetization vector at the surface \cite{Rado59,Gurevich96,Guslienko05}, regardless of the microscopic sources of SW pinning pertaining to the changes of physical and chemical states at the surface. This pinning mechanism, called here {\it exchange pinning} only because of its local character, is dominant in magnetic structures of small sizes or in thin magnetic layers, where the long-range  dipolar interactions (induced by the presence of the surfaces) cannot compete with the exchange interactions. The exchange pinning was also extensively investigated in lattice models.\cite{ Puszkarski79, Mamica1998, Mamica2015} However, the typical in-plane sizes of the planar magnonic structures investigated experimentally are usually larger than  tens of nanometers and the non-local demagnetizing field is imperative. The unavoidable {\it dipolar pinning}\cite{Guslienko02,Guslienko05,Wang19} is related to the geometry of magnetic structure and the presence of magnetic surface and volume magnetic charges in confined geometries.  

Unfortunately, the consideration of non-local dynamical demagnetizing effects is computationally challenging for such large structures.\cite{Zivieri2005, Mamica2013, Mamica2014, Hussain2018} This makes models assuming a continuous distribution of the magnetization more suitable to include the combined impact of the pinning, dipolar and exchange interactions on SWs both in terms of computational efficiency and theoretical analysis.\cite{Guslienko02, Demokritov03, Guslienko05, Rychly17,Wang2018}

The concept of SW pinning at the edges of the ferromagnetic stripes and holes has also been used to interpret SW spectra in MCs. The pinning of dipolar origin has been exploited in analysis of the SW spectra in single-component MCs. \cite{kostylev2004,Gubbiotti2005} The influence of the surface anisotropy on the SW dynamics has been studied in  magnonic waveguides based on antidot lattice. \cite{Klos12,klos2013} A detailed study of the interplay between the SW pinning and coupling strengths at the interface between two ferromagnetic materials on the spectrum of MCs can be found in Ref.~[\onlinecite{Kruglyak2017}].

The starting point of our investigation are the results of Brillouin light scattering measurements of the SW dispersion relation in an 1D MC composed of Co and Py stripes presented in Ref.~[\onlinecite{Wang09}]. In this paper, the existence of magnonic band gaps in a bi-component MC was experimentally demonstrated. This paper has become a point of reference for a number of further theoretical and experimental investigations where different aspects of SW dynamics in MCs were analyzed. Nevertheless, the influence of a thin nonmagnetic layer separating two metallic ferromagnetic stripes in bi-component MCs on the SWs spectra has not yet been investigated.

In our study we investigated how the structural parameters of single- and bi-component 1D MC (see Fig.~\ref{Fig:structure}) affect the  SW pinning at the stripes' edges. We checked how the width of the spacer between the stripes in the periodic sequence influences the low frequency SW mode with in-phase oscillations of the magnetization in the whole structure, i.e., at the ferromagnetic resonance (FMR), also called a fundamental mode. We  focused on the structures most readily accessible for experimental investigations, i.e., those in which the magneto-dipolar coupling and pinning dominate over the exchange interaction. Therefore, we  investigated the SW dynamics using the model where the magnetic structures are described by the spatial distribution of the magnetic parameters: the saturation magnetization and exchange stiffness constant. We considered the pinning which depends on the stripe geometry (dipolar pinning) and the SW pinning resulting from the surface anisotropy at the stripes' edges (exchange pinning). We showed that both are important factors that alter the SW dispersion, modify the dynamical coupling between the stripes, and influence the SW propagation. Our theoretical and numerical studies of SWs in such structures allowed us to shed light on this unexplored area.

We employed three numerical methods -- the plane wave method (PWM), a finite element method (FEM), and a finite difference time domain (FDTD) method -- to calculate the SW spectra in the frequency and time domain. These methods are fully described in the next section (Sec.~\ref{model}). Then, we studied the influence of the separation between the stripes of the same material in single-component MC on SW spectra and SW pinning (Sec.~\ref{Sec:Single}). In Sec.~\ref{Sec:Bicomponent} we presented the dynamical coupling in bi-component MCs, focusing on their influence on the FMR frequency, formation of the magnonic bands and their dependence on the magnetic field. The paper is ended with Sec.~\ref{Sec:Conclusions} where we summarised our results.

\section{Theoretical modeling\label{model}}
\begin{figure}[!ht]
\includegraphics[width=8cm]{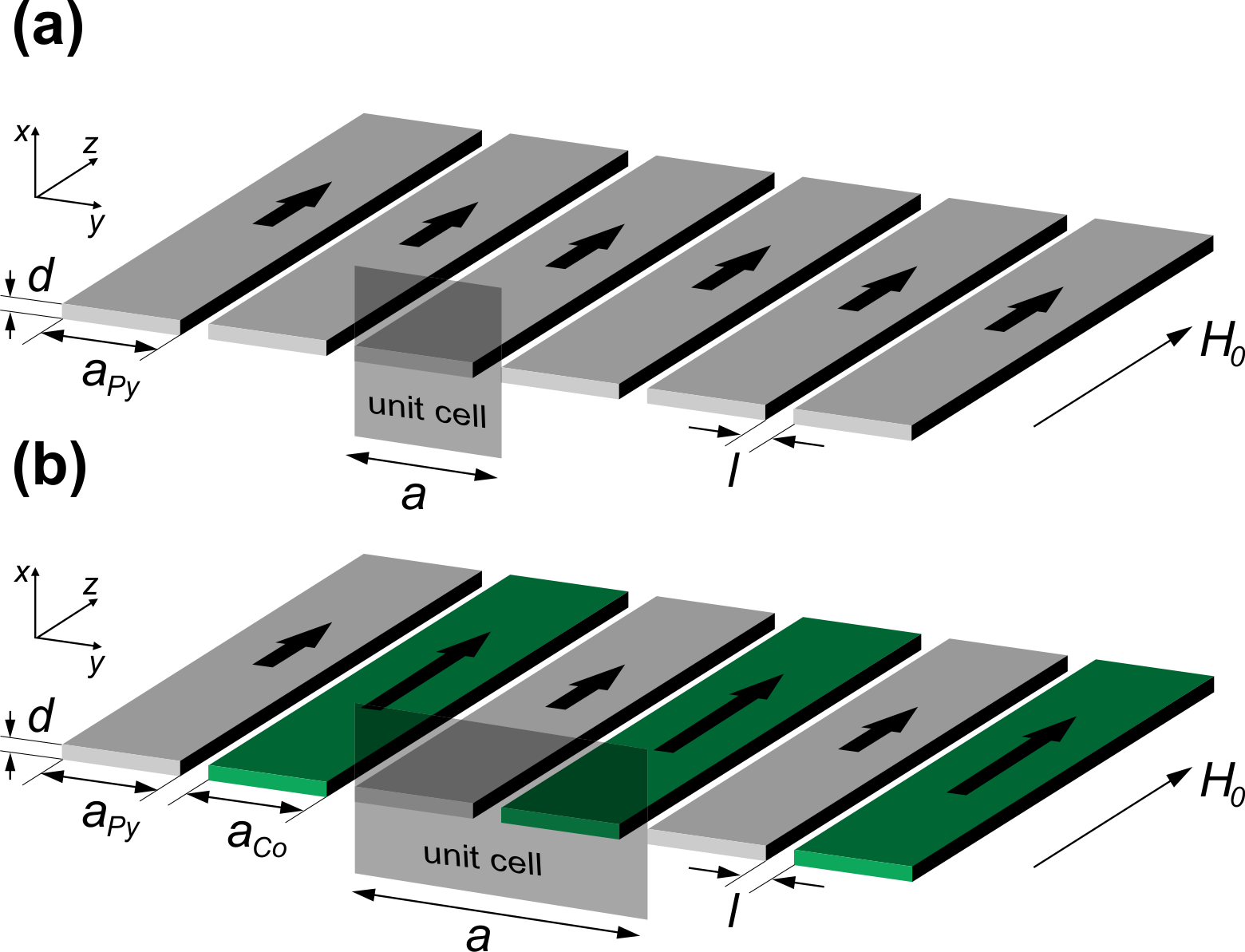}
   \caption{(a) \small{Geometry of a single-component magnonic crystal (MC) in the form of an 1D array of Py stripes, where the stripes' width is $a_{\text{Py}} = 250$ nm, and the lattice constant is $a = a_{\text{Py}} + l$.} (b) \small{Geometry of a bi-component MC in the form of an 1D periodic array of Co and Py stripes, where the stripes' width is universally $a_{\text{Py}} = a_{\text{Co}} = 250$ nm, and the lattice constant is $a = a_{\text{Py}} +  a_{\text{Co}} + 2l$. In both (a) and (b), stripe thickness is $d$, and the separation between the stripes' by a non-magnetic spacers (NMSs)  is $l$. The stripes are magnetically saturated along their axes. Spin waves (SW) propagate within the plane of the MC, perpendicular to the magnetization. The unit cell used in  calculations (periodically repeated along the $y$ axis) is marked by the semi-transparent box.}}
\label{Fig:structure}
\end{figure}

We consider two types of 1D MCs. The first one is a single-component MC composed of an 1D periodic array of Py nanostripes of equal width $a_{\text{Py}} = 250$ nm [Fig.~\ref{Fig:structure}(a)], and the second is a bi-component MC composed of alternating Py and Co stripes of equal width $a_{\text{Py}} = a_{\text{Co}} = 250$ nm [Fig.~\ref{Fig:structure}(b)]. The stripes have thickness $d$, infinite length (along the $z$ axis) and are either in touch or separated by air (or by any other dielectric non-magnetic spacers  (NMSs) of width $l$.  The MC is assumed to be magnetically saturated along the $z$-axis, even when the static external magnetic field $H_0$ (pointing in the same direction) is set to $H_0 = 0$.\cite{Adeyeye:1639}

The assumption, that the magnetization is in its equilibrium configuration allows us to use the linear approximation in SW calculations, which implies small deviations of the magnetization vector ${\mathbf{M}}({\mathbf{r}},t)$ from its equilibrium orientation. Thus, the magnetization vector can be split into its static and dynamic parts ${\mathbf{M}}({\mathbf{r}},t)=M_{z}({\mathbf{r}})\mathbf{\hat{z}}+{\mathbf{m}}({\mathbf{r}},t)$, and we can neglect all nonlinear terms in the dynamical components of the magnetization vector $\mathbf{m}({\mathbf{r}},t)$ in the equation of motion defined below. Since $|{\mathbf{m}}({\mathbf{r}},t)|\ll |M_{z}(y)|$, we can assume also $M_{z}(y)\approx M_{\text{S}}(y)$, where $M_{\text{S}}(y)$ is the saturation magnetization dependent on the $y$-coordinate. We consider here monochromatic SWs in fundamental mode ($\mathbf{k}=0$) which give the main contribution to the FMR signal. To investigate the impact of NMSs on the SW band structure, additionally consider that the SWs are propagating along the direction of periodicity ($\mathbf{k}=k_y \hat{\mathbf{y}}$), which are more affected by periodic modulation of the structure than SWs propagating in an oblique direction\cite{Kostylev08b}. In both cases the phase and amplitude of SWs are homogeneous in the $z$-direction ($k_z=0$). Therefore, we can reduce the problem to the $x-$ and $y-$dimension, and write ${\mathbf{m}}({\mathbf{r}},t)={\mathbf{m}}(x,y) \exp(i\omega t)$, where $\omega$ is the
SW angular frequency, $\omega = 2 \pi f$, and $f$ is the frequency. The dynamics of the magnetization vector $\mathbf{m}(x,y,t)$ with negligible damping is described by the stationary Landau-Lifshitz equation:
\begin{eqnarray}
i \omega {\mathbf{m}}(x,y)
=-|\gamma|\mu_{0} \left( M_{\text{S}}(y)\mathbf{\hat{z}}+{\mathbf{m}}(x,y) \right) \times{\mathbf{H}}_{\text{eff}}({\mathbf{r}}),
\label{eq:LL}
\end{eqnarray}
where $\gamma$ is the gyromagnetic ratio, $\mu_{0}$ is the permeability of vacuum, and $\mathbf{H}_{\text{eff}}$ denotes the effective magnetic field acting on the magnetization. 

The effective magnetic field $\mathbf{H}_{\text{eff}}$ is in general the sum of several components. Here, we will consider four terms:
\begin{equation}
\mathbf{H}_{\text{eff}}(\mathbf{r},t)=H_{0}\hat{\mathbf{z}}+\mathbf{H}_{\text{ex}}(\mathbf{r},t)+\mathbf{H}_{\text{dm}}(\mathbf{r},t) + H_{\text{ani}} \hat{\mathbf{z}}.\label{eq:Heff}
\end{equation}
The second term is the exchange field, assumed to have the form:
\begin{equation}
\mathbf{H}_{\text{ex}}(\mathbf{r},t) = \nabla \cdot \left( \frac{2A(y)}{\mu_0 M_{\text{S}}^2(y) }\nabla \mathbf{m}(\mathbf{r},t) \right),
\end{equation}
where $A$ is the exchange stiffness constant.\footnote{We have taken here the exchange field in the form used in the previous papers dedicated to Co/Py bi-component MC.\cite{Wang09} It is different from the form used in our previous paper Ref.~[\onlinecite{Sokolovskyy11}] but we have verified that the dependence is the same, while differences between these two formulations for the stripes of 250 nm size are less than 4\% with respect to the frequency.\cite{Krawczyk12}}
The third term of the effective field is the demagnetizing field. In the linear approximation, in a similar manner to the magnetization vector, it can be decomposed into its static and dynamic components,   $\mathbf{H}_{\text{dm}}(\mathbf{r})$ and $\mathbf{h}_{\text{dm}}(\mathbf{r},t)$, respectively. In the case of infinitely long stripes saturated along their axis, the static demagnetizing field along this axis vanishes. The time and space dependence of the dynamic component of the demagnetizing field is assumed to have the form $\mathbf{h}_{\text{dm}}(\mathbf{r},t)=\mathbf{h}_{\text{dm}}(x,y) \exp(i\omega t)$. This can be calculated in either the reciprocal or real space from Maxwell's equations within the magnetostatic approximation\cite{Gurevich96}, and then implemented in Eq.~(\ref{eq:LL}) \cite{Sokolovskyy11,Mruczkiewicz13}.  

The magnetic parameters $A$ and $M_{\text{S}}$ are periodic functions of the position along the $y$-axis, with a period $a=a_{\text{Py}}+l$ or $a=a_{\text{Py}}+a_{\text{Co}}+2l$ for single-component and bi-component MCs, respectively. Thus, the Bloch theorem holds, asserting that a solution can be represented as a product of a plane wave envelope function and a periodic function ${\tilde{\mathbf{m}}}_{k_y}(x,y)$ along $y$ with the period $a$:
\begin{equation}
{\mathbf{m}}(x,y)={\tilde{\mathbf{m}}}_{k_y}(x,y) e^{i k_y y},
\label{eq:Bloch}
\end{equation}
where $k_y$ is the Bloch wave vector of SWs propagating along the $y$-axis, which can be limited to the first Brillouin zone. To calculate the magnonic band structure and amplitude of SWs in the considered MCs, we employed the PWM and FEM to solve the eigenproblem obtained from Eq.~(\ref{eq:LL}) with the use of Bloch theorem Eq.~(\ref{eq:Bloch}). 

In the FEM, the equations are solved on a two-dimensional discrete mesh in real space (in the plane ($x$, $y$)) limited due to the Bloch theorem to the single unit cell [marked by the gray box in Fig.~\ref{Fig:structure}]. In this paper, we use one of the realizations of the FEM developed in the commercial software COMSOL Multiphysics ver. 4.2. This method has already been used in calculations of magnonic band structure in thin 1D MCs, and their results have been validated by comparing with micromagnetic simulations and experimental data.\cite{Wang09,Chi11,Mruczkiewicz13} The detailed description of the FEM in application to calculations of SW spectra in MCs can be found in Refs.~[\onlinecite{Mruczkiewicz13,Mruczkiewicz.2013b}]. The exchange pinning was introduced in the FEM calculations by  forcing the  boundary conditions  for dynamic component of the magnetization ($m_{y}=0$ and $dm_{x}/dy=0$)\cite{Gurevich96} on the lateral sides of stripes.

In the PWM, all periodic functions of the position: ${\tilde{\mathbf{m}}}_{k_y}(y)$, $M_{\text{S}}(y)$, $A(y)$  in Eqs.~(\ref{eq:LL}) and (\ref{eq:Bloch}) are transformed into reciprocal space using the 1D Fourier transformation, which leads to an algebraic eigenvalue problem.\cite{Sokolovskyy11,Gallardo2018} Then, standard numerical routines are used to find its eigenvalues and eigenvectors. In the PWM, it is necessary to assume uniform magnetic field and magnetization across the sample's thickness. Thus, in the calculations we use the value of ${\bf h}_{\text{dm}}$ from the top surface of the MC. This approach gives correct results for the Damon-Eshbach geometry considered in this paper.\cite{Krawczyk13b} The representation of the dynamic components of the magnetisation vector as a superposition of  plane waves forces one to treat an MC as a continuous medium with eigenfunctions defined at its every point. Thus, it is necessary to define the  artificial magnetic material also for the NMSs, which will not support the existence of SWs in NMSs for considered frequency range. We have shown in Ref.~[\onlinecite{Klos12}] that this can be done by assuming, for the NMSs, small values of $M_{\rm S}$ and a (bulk) uniaxial magnetic anisotropy field parallel to the static comment of magnetization. The small value of $M_{\rm S}$ reduces the amplitude of any nonphysical solutions almost to zero whereas the high anisotropy rises significantly their frequencies. As a result, any spurious solutions found in NMSs have the frequencies far above the considered range and do not affect the modes in magnetic material. However, the side effect of this approach is that large bulk anisotropy (in the nonmagnetic material) can pin the can pin the SWs at the lateral interfaces of the magnetic stripes. Therefore, by changing the value of the artificial magnetic bulk anisotropy in NMSs, we can tune the pinning of SWs reaching the limits of fully pinned and almost unpinned SWs (with nonphysical solutions pushed to much higher frequencies). The control of the spin wave pinning in PWM is difficult, particularly in the regime of weak pinning. In this method, the actual value of surface anisotropy $K_S$ cannot be precisely determined. It can be done only indirectly by comparing the outcomes of PWM with the results of the other methods.

In the FDTD, the considered system is discretized using regular cuboids of fixed size, and then the full Landau-Lifshitz equation (without linearization) is numerically solved across the mesh, assuming a uniform magnetization and effective field within each cell. We employed Object-Oriented Micro-Magnetic Framework (OOMMF)\cite{oommf}, with the implementation of either one\cite{Lebecki} or two-dimensional\cite{Wang_pbc} periodic boundary conditions in order to model an isolated stripe or an MC, respectively. The FDTD calculations were performed in the absence of exchange pinning.

In all mentioned techniques, the dipolar pinning results from the presence of the dynamic demagnetising fields which are included directly in these calculations. Therefore, we are not introducing the dipolar pinning explicitly.

Formally, the strength of spin wave pinning is strictly related to the boundary conditions for dynamical components of the magnetization vector\cite{Rado59,Guslienko05}. The energy density of precessing magnetization can be different at the surface due to various local effects (expressed by the surface anisotropy $K_S$) or because of the magnetic charges induced by the presence of the surface (resulting from non-local dipolar interactions\cite{Ivanov02}). This surface density of energy can be related to the effective field and the additional torque which acts on the magnetization across the surface. The surface torque $\mathbf{T}_{\rm surf}$  must be compensated by the torque related to the truncation of the exchange interaction at the surface. The boundary conditions for magnetization, called Rado-Weertman boundary conditions\cite{Rado59}, can be derived from the balance of these surface densities of torques:
\begin{equation}
    \mathbf{M}\times\mu_0l_{\rm ex}^2 \frac{\partial\mathbf{M}}{\partial n}+\mathbf{T}_{\rm surf}=0,
\end{equation}
where the derivative $\partial/\partial n$ is taken along the unit vector $\mathbf{n}$  normal to the surface. The parameter $l_{{\rm ex}} = \sqrt{\frac{2A}{\mu_{0}M^2_{S}}}$ denotes the exchange length. In the absence of the torque $\mathbf{T}_{\rm surf}$, the exchange interactions prefer to unpin the magnetization on the surface $\frac{\partial\mathbf{M}}{\partial n}=0$. The torque $\mathbf{T}_{\rm surf}$ forces the reorientation of the magnetization (or changes the amplitude of magnetization dynamics) close to the surface which induces the volume charges and increases the exchange energy at the cost of the surface charges and surface energy density. This process results in the magnetization pinning. Thus, the magnetization pinning is caused both by the local effects (existing even in the absence of dipolar interaction) and long-range dipolar interactions. We call these mechanisms exchange pinning and dipolar pinning respectively.

The Rado-Weertman boundary conditions for the dynamic component of the magnetization normal to the lateral faces of the $i$-th stripe can be expressed with
\begin{equation}
    \left.a_i\frac{d m_y(x,y)}{d y}\mp p\, m_y(x,y)\right|_{y=y_+,y_-} =0,
\end{equation}
where $a_i=a_{\rm Py},a_{\rm Co}$ is the width of the $i$-th stripe and $y_+$, $y_-$ are the positions of the right and left edges of the stripe ($y_+>y_-$). The symbol $p$ denotes the dimensionless pinning parameter, where the limiting cases $p=0$ $\left(\frac{d m_y}{d y}=0\right)$ and $p\rightarrow\infty$ $\left(m_y=0\right)$ correspond to completely free and totaly pinned dymamic comonet $m_y$ at the lateral edges $y=y_+,y_-$, respectively. For the isolated stripe of small aspect ratio ($d/a_i\ll 1$) the parameter $p$ may be approximated as \cite{Guslienko05,Guslienko02}:
\begin{equation}
   p=\frac{2\pi\left(1-\frac{4 K_{S,i}}{\mu_0 M_{S,i}^2 d}\right)}{\frac{d}{a_i}\left(1+2\ln{\left(\frac{a_i}{d}\right)+\left(\frac{l_{{\rm ex},i}}{d}\right)^2}\right)}. \label{eq:pinning}
\end{equation}
 The symbol $K_{S,i}$ represents the surface anisotropy with respect to the direction normal to the lateral faces of the stripe: $\mathbf{n}=\pm \hat{\mathbf{y}}$. 
The large and positive value of $K_{S,i}\gg \mu_0 M_S^2 d/4$ results in the strong exchange pinning of SWs.  In the absence of  surface anisotropy ($K_{S,i}=0$) the SW pinning (\ref{eq:pinning}) is determined  by the dipolar interactions only. Since the dipolar pinning for a magnetic stripe is determined by the geometrical parameters (width $a_i$ and thickness $d$), it is significantly reduced when the stripes are downsized below the exchange length: $\sqrt{d\;a_i}\ll l_{{\rm ex},i}$\cite{Wang19}. When the dimensions of the stripe are larger than exchange length,  the dipolar pinning is enhanced with the increase of the ratio: $a_i/d$. 

The dipolar pinning is  further changed by the dipolar interaction between the stripes, which depends in our structures on the width of the NMSs $l$. When the stripes merge into continuous magnetic layer ($l\rightarrow 0$), the spin wave pinning will disappear and we will observe the increase of the FMR frequency to the FMR frequency of continuous layer. This suggests that by the reducing of $l$
Therefore, the change of the SW pinning by the adjustment of geometrical parameters is an important factor for shaping the spectrum of considered periodic structures.

In our numerical model [based on Eq.~(\ref{eq:LL}) and (\ref{eq:Heff})] the dipolar pinning is directly included via the dynamic demagnetizing field $\mathbf{h}_{\rm dm}(\mathbf{r},t)$. Therefore, the effect of the dipolar pinning is reproduced in the outcomes of numerical methods PWM, FEM and FDTD. However, the exchange pinning resulting from nonzero surface anisotropy ($K_{S,i} \ne 0$) is an interface effect and has been introduced manually. We focus here on two limiting cases: (i) $K_{S,i}=0$, when the SW pinning is governed only by dipolar effects, and (ii)  full exchange pinning, when the SWs are completely pinned at the lateral faces of the stripes ($K_{S,i}\gg\mu_0 M_{S,i}^2 d/4$) and the dipolar effects can affect the amplitude of SW inside the stripes only.

\section{Results and discussion \label{res}}
\subsection{Single-component magnonic crystals}\label{Sec:Single}
\begin{figure}[!ht]
\includegraphics[width=8cm]{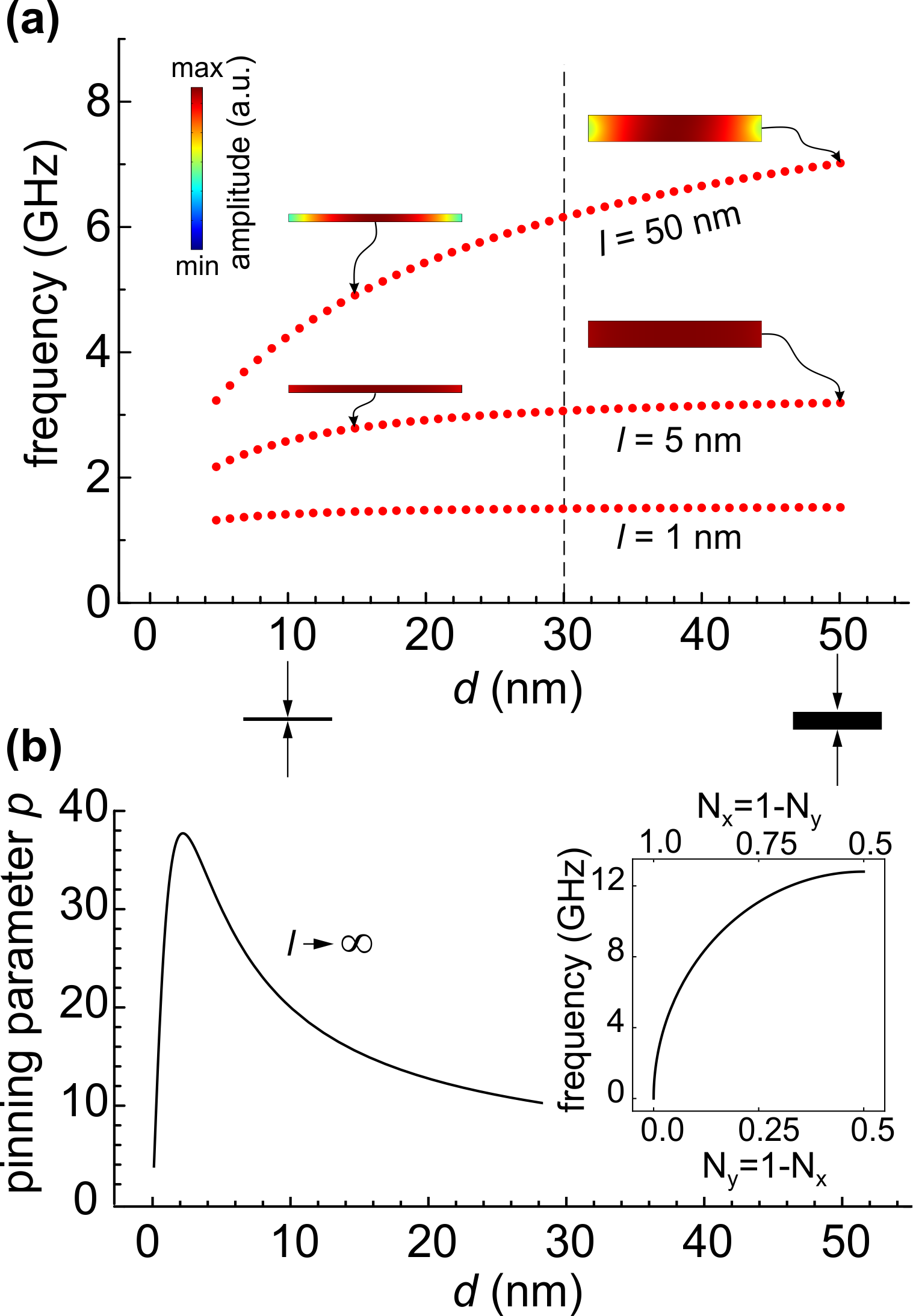}
   \caption{\small{(a) The dependence of the frequency of fundamental SW mode ($k_{y}=0$) on the thickness $d$ of Py stripes in a single-component MC. The width of the NMSs between the stripes $l$ is fixed to $1, 5$ or $50$ nm. The width of the stripes $a_{\rm Py}$ is equal to 250 nm. The calculation was done using the FEM for $H_0 = 0$ in the absence of the surface anisotropy (i.e., absence of exchange pinning). To illustrate the impact of the thickness $d$ on the SW pinning, we show selected profiles of the in-plane component of the SW amplitude $m_{\rm y}$ in the cross-section of one stripe.  (b) The  pinning parameter $p$ for $K_s=0$ (in the limit of an isolated stripe: $l\rightarrow\infty$) as a function of the thickness $d$. In the realistic range of $d$, considered in (a), an increase of $d$ makes the SW gradually unpinned, as also illustrated in the SW profiles shown in (a). Dashed vertical line in (a) marks the value of $d = 30$ nm for which the further results in Fig.~\ref{Fig:gap_width_dep} are shown. The inset in (b) presents the dependence of the frequency of fundamental mode on demagnetizing factors. } }
\label{Fig:thickness_dep}
\end{figure}
In this section, we will study the SW spectra in an array of stripes made of permalloy (Py) only [Fig.~\ref{Fig:structure}(a)].\cite{Wang09, Gubbiotti10, Zhang16, Huajun18, Lisiecki19} We assume Py to have standardized values of spontaneous magnetization ($M_{\text{S},\text{Py}} = 658$ kA/m) and exchange constant ($A_{\text{Py}} = 11$ pJ/m) as obtained experimentally in Ref.~[\onlinecite{Wang09}]. The gyromagnetic ratio is assumed to be $|\gamma| = 1.946\times 10^{11}$ (rad/s)/T. The exchange length of Py is  $l_{\rm{ex,Py}} =~6.4$~nm, and so, the cell size adopted in FDTD calculations is set to (30$\times$5$\times$5)~nm$^3$.

\begin{figure}[!ht]
   \includegraphics[width=8cm]{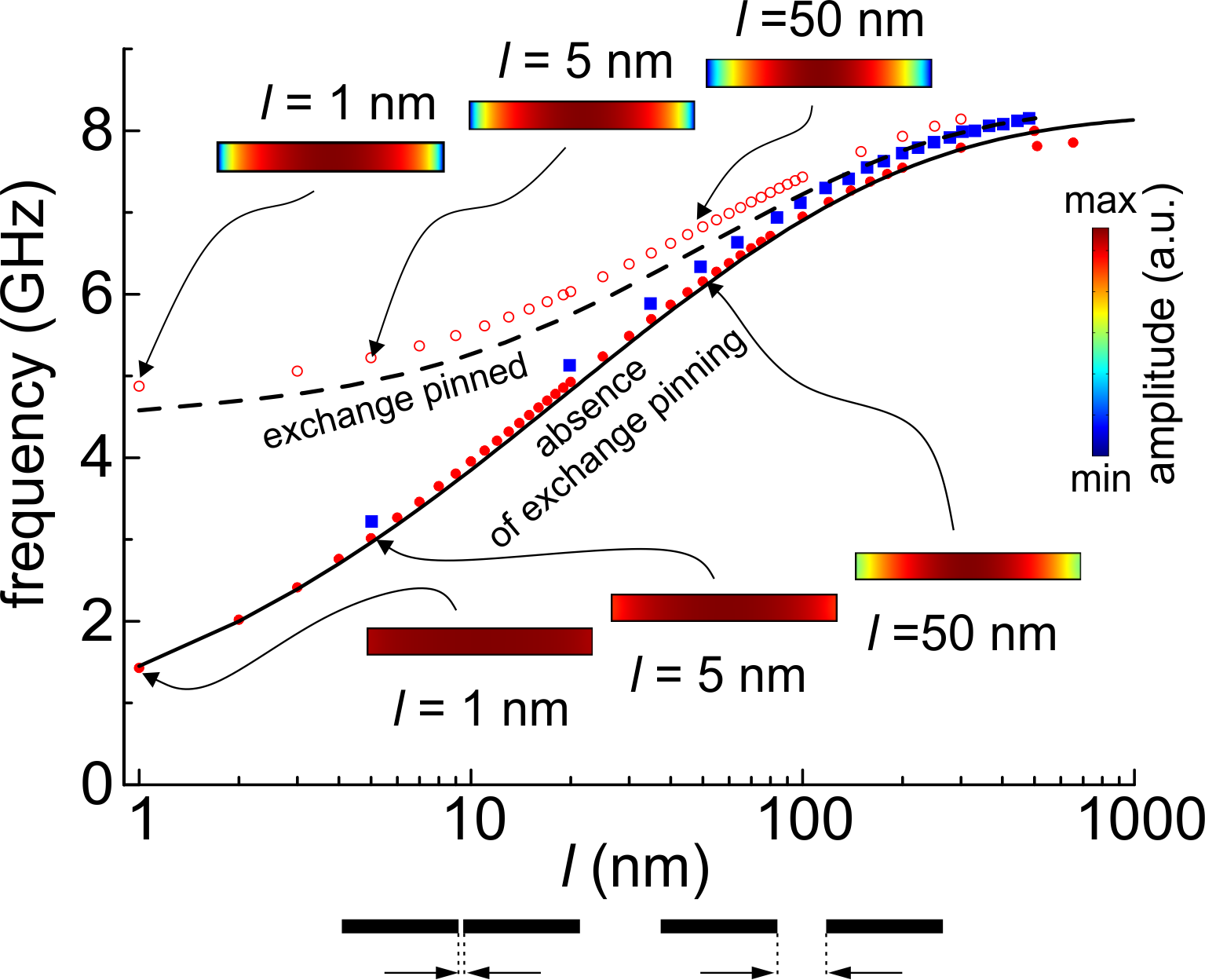}
   \caption{\small{The dependence of the frequency of the fundamental SW mode ($k_{y} = 0$) on the separation between Py stripes, $l$, for $H_0 = 0$. The stripe separation is scaled logarithmically. The red full and empty dots are the results of FEM calculations, the blue squares are the results of FDTD calculations, and the lines correspond to the PWM results for two limiting cases which concern the SW pinning at the edges of stripes: absence of exchange pinning ($K_{S,{\rm Py}}=0$) -- solid line and strong exchange pinning ($K_{S,{\rm Py}}\gg\mu_0 M_{S,{\rm Py}}^2 d/4$) -- dashed line. The insets show the distribution of the in-plane component of SW amplitude  $m_{\rm y}$ in the cross-section of one stripe for $l =$ 1, 5 and 50 nm. The calculations were done for the stripes' thickness $d$ and width $a_{\rm Py}$  equal to  30~nm and 250~nm, respectively.}}
\label{Fig:gap_width_dep}
\end{figure}

Let us begin by considering the impact of the dipolar pinning on the SW spectrum. Hence, initially, we assume $K_{S,{\rm Py}}=0$. In Fig.~\ref{Fig:thickness_dep}(a) we present the frequency of the fundamental SW mode as a function of the stripe thickness $d$ for three MCs differing in the width of NMSs  ($l=50$, 5 and 1 nm), calculated by the FEM. For the  widest air gap the FMR frequency $f$ increases monotonously with $d$, while for small $l$ the frequency $f$ is almost independent of $d$, as expected for a continuous magnetic film. 
By inspection of Eq.~(\ref{eq:pinning}), where the influence of the $y$ component of the dynamical magnetization  on SW pinning is only taken into account, we can notice that the increase of thickness initially enhances the dipolar pinning $p$ (i.e., for $d$ smaller or comparable to the exchange length $l_{\rm ex,Py}$), while for larger thicknesses [as shown in Fig.~\ref{Fig:thickness_dep}(a)] it slowly reduces the value of $p$ with increasing $d$ -- see Fig.~\ref{Fig:thickness_dep}(b), that shows $p(d)$ for an isolated stripe. This effect is also visible in the SW profiles in stripes comprising an MC, which are presented in the insets of Fig.~\ref{Fig:thickness_dep}(a). We expect intuitively that the weakening of the SW pinning should result in a downward frequency shift because of the weaker confinement of the SW modes within the relatively narrow stripes. However, the trend observed in Fig.~\ref{Fig:thickness_dep}(a) is opposite. It can be explained using the Kittel formula for the FMR frequency for an isolated stripe:\cite{Gurevich96}  
\begin{equation}
\omega^2 = \left[\omega_0 +(N_x -N_z) \omega_{\text{M}}\right] \left[\omega_0 +(N_y -N_z) \omega_{\text{M}} \right], \label{Eq:Kittel}
\end{equation} 
where $\omega_0 = |\gamma| \mu_0 H_0$, $\omega_{\text{M}} = |\gamma| \mu_0 M_{\text{S,{\rm Py}}}$ and $N_x$, $N_y$ and $N_z$ are the effective demagnetizing factors with respect to the out-of-plane direction, along the periodicity direction and along the external magnetic field, respectively. For the infinitely long stripes, $N_z=0$ and $N_y=1-N_x$. The gradual increase of the thickness $d$ from 0 to $a_i$ results in the change of the demagnetizing factors in the range $0\rightarrow 0.5$ for $N_y$ and $1\rightarrow 0.5$ for $N_x$. From the Kittel formula (\ref{Eq:Kittel}), we find that increasing the thickness of an initially thin stripe ($N_y\ll N_x$) results in a rapid increase of the FMR frequency, as shown in the inset of Fig.~\ref{Fig:thickness_dep}(b). This understanding can be transferred to the MC composed of the dipolar interacting stripes shown in Fig.~\ref{Fig:thickness_dep}(a) for $l=50$ nm. However, the stronger dipolar interactions between stripes (as observed in Fig.~\ref{Fig:thickness_dep}(a) for narrow air gaps: $l=1$ or 5~nm) reduces both the dipolar pinning (see SW profiles) and the FMR frequency. The observed flattening of the $f(d)$ dependence with decreasing $l$, approaching the case where $f$ is close to independent of $d$ for $l=1$ nm and $d > 5$ nm, can be explained when we notice that at FMR frequency (i.e., the frequency of fundamental mode at  $k_y$=0) the considered system reaches the metamaterial limit.\cite{Mikhaylovskiy10,Mruczkiewicz2012} In this limit the 1D MC with small air gaps can be treated as a homogeneous thin film with  effective parameters characterizing the anisotropy and magnetization saturation.  


To explain fully the effect of a dynamic coupling between the stripes, we performed  systematic numerical simulations of the FMR frequency (i.e. the frequency of fundamental mode) $f$ as a function of $l$ (the width of the NMSs). The $f(l)$ dependencies calculated in the range 1~nm~$<l< 1$~$\mu$m with $H_0 = 0$, are presented in Fig.~\ref{Fig:gap_width_dep}. The data showing FMR frequencies indicated by the red-filled dots, blue-filled squares and solid black line, were calculated using FEM, FDTD and PWM methods respectively, in the absence of surface anisotropy on the lateral faces of  the Py stripes ($K_{S,{\rm Py}} = 0$), where the exchange SW pinning is not active. The set of results marked by empty dots and dashed lines were calculated using FEM and FDTD calculations respectively, for the alternative limiting case in which there is strong surface anisotropy ($K_{S,{\rm Py}}\gg\mu_0 M_{S,{\rm Py}}^2 d/4$) i.e., the SWs are exchange pinned.

 Before providing physical interpretation of the results, we compare the results of the FEM, FDTD and PWM methods. The results obtained using the FDTD  (blue-filled squares) are in excellent agreement with the results obtained from the FEM (red-filled dots)  when no exchange pinning is included. In the PWM calculations, the SW pinning appears due to the high value of the fictitious anisotropy field used in NMSs. To avoid introduction of this artificial SW pinning in the PWM, we reduced the bulk anisotropy field inside the NMSs, which guarantees the almost full unpinning of fundamental mode. By artificially inserting a bulk anisotropy field of strength 0.7 T within the NMSs, we obtain excellent agreement between the FMR frequencies calculated using PWM with the FMR frequencies obtained with the aid of FEM and FDTD method. This agreement is sustained as well for larger separation $l$  (the difference between PWM and FEM calculations is only 0.15 GHz at $l = 900 $~nm). 
  In the case of strong exchange pinning, we can easily obtain the corresponding result from FEM (red-open dots) and PWM (black dashed line) calculations. To find the effect of the dynamic magnetization pinning in the PWM, we have to use the significantly higher value of anisotropy field for NMSs.\cite{Klos12,Kumar14} While this increase does not increase the SW pinning (as seen by the lack of change in the FMR frequency), the anisotropy can impact the convergence of the PWM. The differences between FEM and PWM outcomes become 0.29 and 0.15 GHz at $l=1$ and 900~nm, respectively, and are related to the assumptions made in the PWM (i.e., uniform dynamics across the thickness and demagnetizing field taken from the top surface of MC).

For both boundary conditions (Fig.~\ref{Fig:gap_width_dep}) we observe a  substantial increase of the FMR frequency upon increasing the width of the NMSs, in accordance with previous studies.\cite{kostylev2004} The frequency increases up to 8.12 GHz (8.44~GHz in the calculations with a pinned dynamic magnetization) at $l = 900$~nm. Interestingly, the  saturation of the dependence $f(l)$ is approached when the stripe separation l becomes larger than the stripes' width $a_{i}$. The $f(l)$ dependence  spans the range of values between two limits for which analytical solutions are known. These are $l=0$ and $l \rightarrow \infty$, i.e., a thin uniform ferromagnetic film\cite{Stancil09, Kalinikos86} and an isolated stripe.\cite{Guslienko02} The FMR frequencies in these limits can be estimated from the Kittel formula (\ref{Eq:Kittel}).  For a thin film ($l = 0$) the demagnetizing field is uniform, and so $N_x = 1$ and $N_y = N_z = 0$. With $H_0 = 0$ this gives $\omega = 0$. However, the demagnetizing field is nonuniform in a finite ferromagnetic body of  non-ellipsoidal shape, and so the demagnetizing factors in general become spatially-varying in general.\cite{Joseph65} Nevertheless, in typical interpretations of experimental results, the FMR frequency for stripes is estimated using fixed demagnetizing factors within Eq.~(\ref{Eq:Kittel}).
\cite{Hoffman1982,Pant1996,Mart_nez_Huerta_2013,Zhu_2018,Lisiecki19} However, the difference between the $f(l)$ dependencies for fully pinned dynamical components of the magnetization and for the absence of  exchange pinning at the stripe edges suggests that neither the dynamical surface charges, nor the pinning parameter $p$, nor constant demagnetization factors are sufficient to describe dynamic dipolar coupling in MCs. The change of the FMR frequency with $l$ for the fully  pinned case points at the non-negligible contribution of the dynamic volume magnetic charges formed due to nonuniform amplitude of SWs inside the stripes. 

\begin{figure}[!ht]
   \includegraphics[width=8cm]{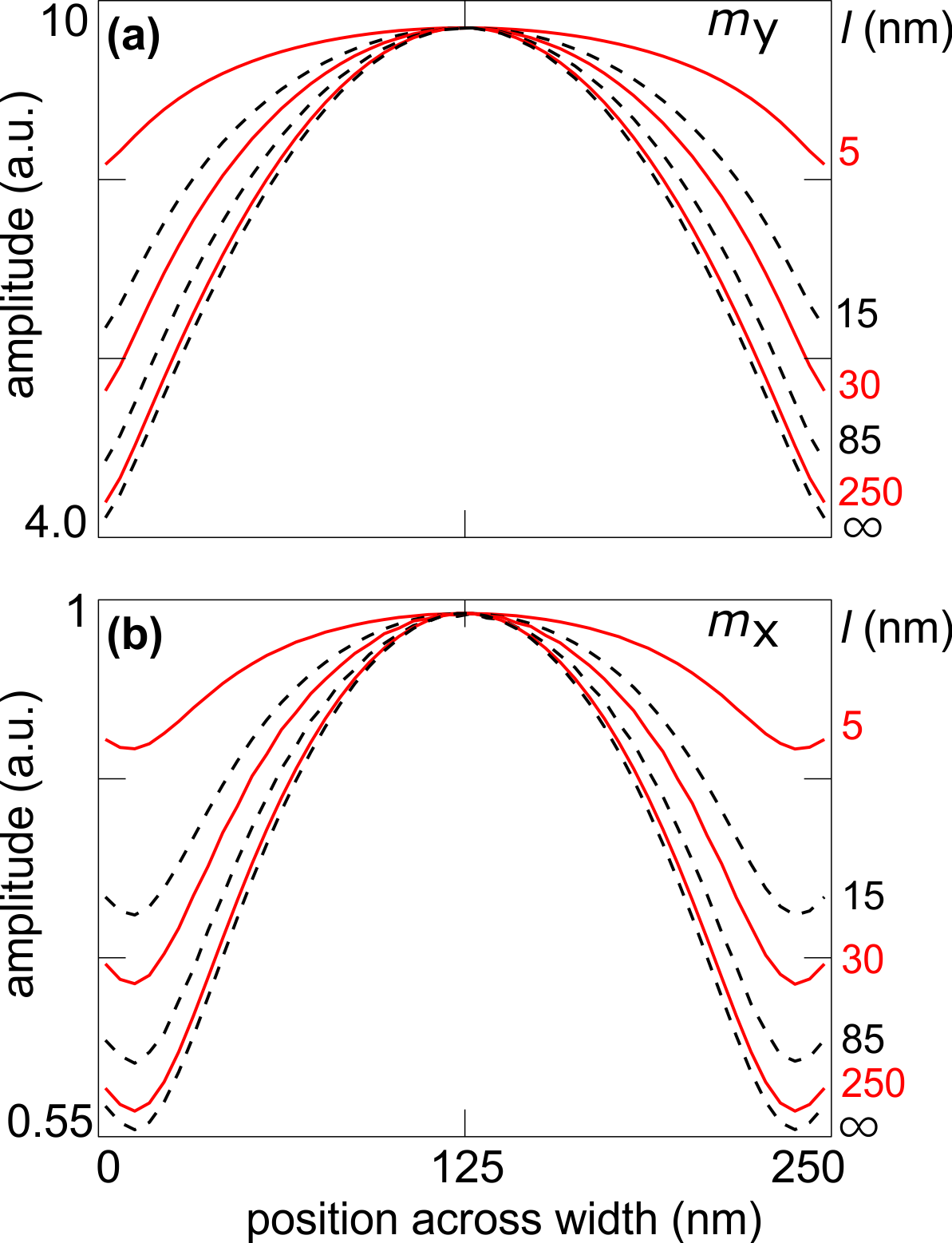}
   \caption{(a)-(b) \small{The amplitude of the $m_{y}$ and $m_{x}$ components of the dynamic magnetizations across the width of the stripe, respectively, calculated at the  FMR  frequency as identified from Fig.~\ref{Fig:gap_width_dep}. These profiles were obtained from the FDTD calculations (in the absence of exchange pinning and for $d =$ 30~nm, $a_{\rm Py} =$ 250 nm). The magnetization profiles shown correspond to different separation $l$ values, as indicated, ranging from $l=$ 5 nm to $l= \infty$. }}
\label{Fig:profiles}
\end{figure}

In order to further explain the dependencies shown in Fig.~\ref{Fig:gap_width_dep}, and in particular to ascertain the relative roles of the dynamic magnetization pinning and the stray fields from neighbouring stripes on the FMR frequency, we calculate the SW amplitude across the width of the stripe at resonance. The results are presented in the inset of Fig.~\ref{Fig:gap_width_dep}, which shows that as the stripe separation increases, the amplitude becomes increasingly concentrated towards the centre of the stripe. The finite width of the stripe leads to a symmetrical non-uniform magnetization profile across the stripe's width. The normalized Fast Fourier Transform of the dynamic $x$ and $y$ components of magnetization in time domain at the FMR frequency are presented in Fig.~\ref{Fig:profiles}(a)-(b), respectively, as a function of the interstripe separation, as calculated from the FDTD.

In the limiting case of $l=$ 0 (and with non-zero $H_0$), the magnetization profile across the infinite continuous film would be uniform at resonance. Upon setting $l=5$~nm (the smallest stripe separation that can be implemented in our FDTD calculations), the mode-amplitude distribution in the plane of the stripe becomes sinusoidal-like [Fig.~\ref{Fig:profiles}(a)], with a characteristic maximum at the centre of the stripe. This arises from the reduced influence of the effective dipolar pinning in this region, and also from the symmetry of the problem. As we move closer to the edges, the influence of the SW pinning increases, leading to  more non-uniform magnetization and a reduced amplitude of precession. Increasing $l$ reduces the influence of the stray fields originating from neighbouring stripes, and so the mode profile tends towards that of an isolated stripe ($l=\infty$).

The amplitude of the out-of-plane magnetization component $m_{x}$ at FMR frequency [Fig.~\ref{Fig:profiles}(b)] has a slightly different profile to that of $m_y$. In particular, we see that the out-of-plane profile has a characteristic ‘ridge’ at the edges of the film. Also, the amplitude of $m_{y}$ is about 10 times larger than the amplitude of $m_{x}$ for considered dimensions of stripes, and thus its influence on the dynamical coupling between the stripes is minor.

The observed features can be understood in terms of the interplay between the lateral quantization of the SW modes and the dynamic magnetization pinning. Due to the finite width of the stripe, it is impossible to excite the $k_{y}=$ 0 (i.e., strictly uniform) mode. Instead, a laterally-pinned Damon-Eshbach $n=1$ mode is excited, where $n$ is the antinode number. This leads to the characteristic bell-shaped profile seen in both $m_{y}$ and $m_{x}$. Close to the edge of the stripe (about 15~nm away), dynamic magnetization pinning begins to dominate in the plane of the film. This inhibits the magnetization oscillation in the plane of the film, and so, $m_{y}$ continues to decrease. In contrast, the local inhibition of oscillation in the plane of the film forces the oscillation of the out-of-plane component of magnetization $m_{x}$ to increase. This feature importantly reveals that the dynamic pinning originating from the edges of the stripe is limited in influence to a lateral range of $\approx\frac{d}{2}$, and so does not play a significant role in the variation of the FMR frequency observed in Fig.~\ref{Fig:gap_width_dep}. We speculate that, in principle, one could attempt to explain this feature in terms of the spatial variation of the local susceptibility tensor\cite{Marchenko_12, Davies_17}, although this is beyond the scope of the present paper.  

The features observed in Fig.~\ref{Fig:profiles} also explain the dependencies of the FMR frequency shown in Fig.~\ref{Fig:gap_width_dep}. With decreasing $l$, the stray magnetic fields originating from all the other neighbouring stripes influence the SW dynamics in the given stripe by changing the internal dynamic magnetic field. The non-uniform magnetization profile originating from the lateral SW pinning is always present but is counter-balanced by these stray fields, which promote a uniform magnetization. By imposing artificially pinned dynamic magnetization at the edges of an isolated stripe, the stray field generated from the stripe (at the edges or volume) is reduced in strength, and hence the FMR frequency of the bulk of the stripe increases slightly. Now, as $l$ decreases, the artificially pinned dynamic magnetization reduces the impact of the uniformity induced by the stray fields, and the magnetization across the stripe width at resonance retains significant non-uniformity. This is apparent in the SW amplitudes shown in the inset of Fig.~\ref{Fig:gap_width_dep} for $l=1$ nm. Thus, the difference in the FMR frequencies, between the case of the fully (exchange) pinned SWs and the  case where there is no artificial (exchange) SW pinning,  increases with decreasing $l$, in accordance with the trend observed in Fig.~\ref{Fig:gap_width_dep}.

It is clear that  the main contribution to the change of the FMR frequency observed in Fig.~\ref{Fig:gap_width_dep} is the lateral SW pinning and an inhomogeneous amplitude distribution of the FMR mode appearing as soon as the separation between Py stripes is introduced. This separation leads to the non-uniformity of the internal magnetic field, due to the stray fields from neighbouring stripes and the lateral quantization of the resonant mode.

We can also draw some conclusions for 2D MCs with dots separated from the matrix material with some NMS.\cite{Ding2013,Choudhury2016,Porwal_2018}  Because the demagnetizing fields along all the principal directions can be comparable due to finite thickness and finite in-plane extension, the demagnetization effects (shape anisotropy) are expected to be much weaker than those in 1D MCs. In contrast,  the dynamical coupling via the stray field and the SW pinning influence, although weak, should still persist. The change of SW pinning due to the dipolar interaction between dots in 2D MC should be then observed even for smaller separation between dots than for the case of stripes in 1D MC. Further studies (beyond the scope of this paper) are required to quantitatively treat this problem.

\subsection{Bi-component magnonic crystals}\label{Sec:Bicomponent}
Bi-component MCs with periodicity in 1D and 2D were investigated theoretically and experimentally in Refs. [\onlinecite{Wang09,Sokolovskyy11,Lin11,Klos12b,Krawczyk13b,Choudhury2016, Rychly2017, Silvani18, Huajun18}]. While many interesting properties have been studied, the impact of the stripe separation and SW pinning at the stripe edges on the SW spectra is not well understood. Here, we study an MC composed of an array of Co and Py stripes [Fig.~\ref{Fig:structure}(b)]. Again, the width of the stripes is 250 nm, and the thickness 30 nm. The spontaneous magnetization and exchange constant of Co and Py are equal to those presented in the experimental paper Ref.~[\onlinecite{Wang09}]. For Co, they are $M_{\text{S},\text{Co}} = 1150$ kA/m and $A_{\text{Co}} = 28.8$ pJ/m and for Py, the parameters are the same as those defined in the previous section. We introduce the NMS  of varied width $l$ between Co and Py stripes.

\begin{figure}[!ht]
   \includegraphics[width=8.5cm]{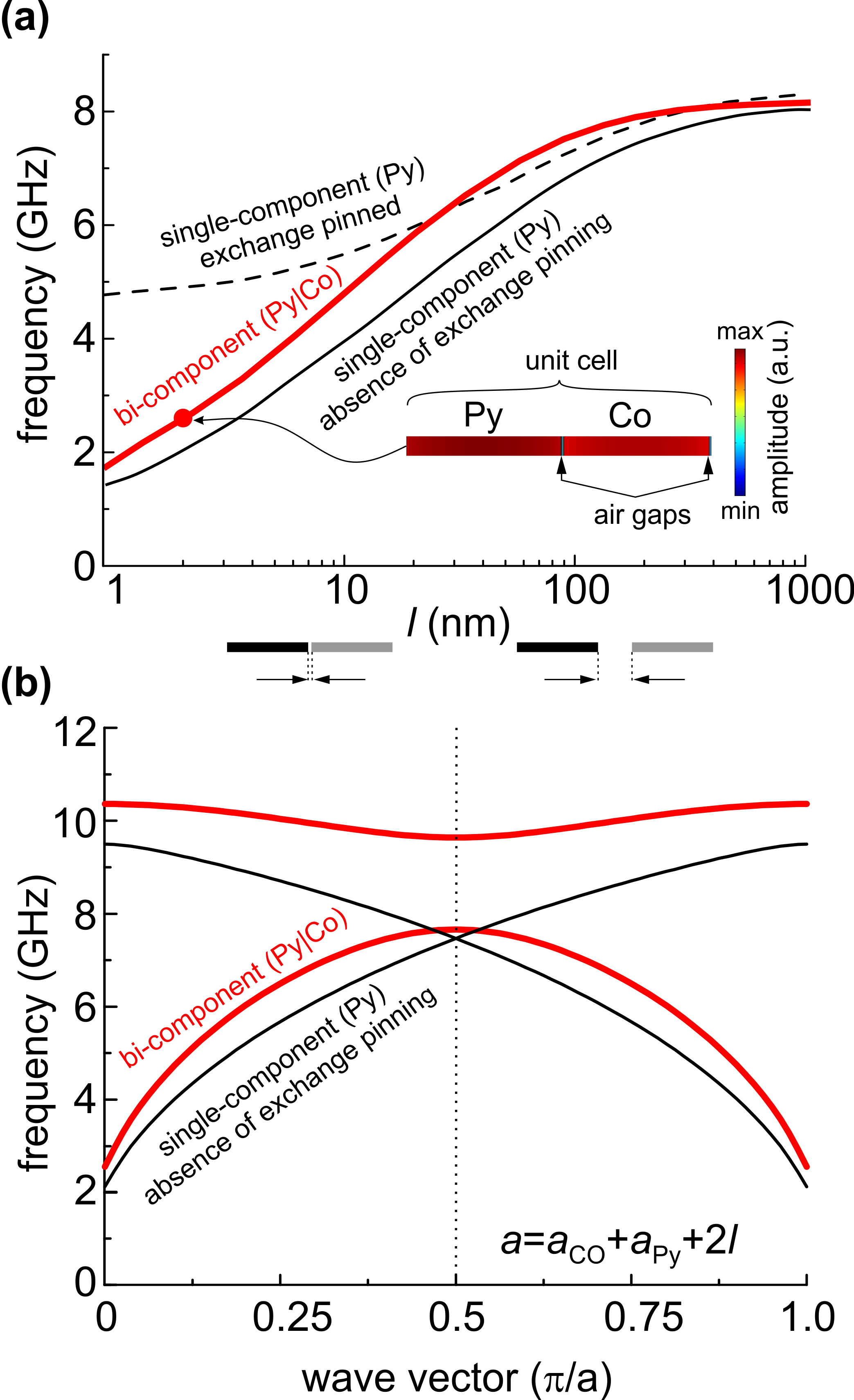}
   \caption{
    \small{(a) The dependence of the frequency of the fundamental mode (FMR frequency) in the 1D bi-component (Co/Py) MC on the separation between the Co and Py stripes, $l$, calculated with the FEM (solid red line). For comparison, the frequencies of the FMR modes of the single component MC composed of Py stripes in the absence of exchange pinning ($K_s=0$) and with exchange pinned SWs at the edges are also shown with dashed black and solid black lines, respectively. The amplitude profile of the fundamental SW mode in the Co/Py MC with a 2 nm-wide NMSs, for $H_0 = 0$, is shown in the inset. (b) The dispersion relations characterizing the two lowest bands for the bi-component MC (red lines) and the lowest band for single-component MC (black line) in the absence of exchange pinning. The dispersion relation for the single-component MC was folded into the first Brillouin zone of the bi-component MC due to our selection of a unit cell which was composed of two Py stripes. We fixed the width of the NMSs to $l=2$ nm. The thickness $d$ and the widths of stripes $a_{\rm Py}=a_{\rm Co}$ are equal to 30 nm and 250 nm, respectivelly, both in (a) and (b).}
   }
\label{Fig:bi_comp_gap_dep}
\end{figure}

With the stripes in contact ($l=0$), the first (fundamental) mode at $H_0=0$ is a quasi-uniform mode spread across the whole MC.\cite{Lin11} The next series of modes form a pattern of standing waves in Py with reduced amplitude in Co.\cite{Wang09,Sokolovskyy11,Lin11} This behavior persists when the NMS is introduced between Py and Co (see mode profile in the inset of Fig.~\ref{Fig:bi_comp_gap_dep}). The dependence of the fundamental mode frequency on the width of the NMS  is shown in Fig.~\ref{Fig:bi_comp_gap_dep} with a thick red solid line. We have used $K_{S,{\rm Py}}=0$ on the Co/Py interfaces here. We observe a dependence similar to that obtained for the single-component MC (for ease of comparison, the solid and dashed lines are taken from Fig.~\ref{Fig:gap_width_dep})). We see that the function $f(l)$ in the bi-component MC (thick red line) merges at large $l$ with the fundamental mode in Py stripes for $K_{S,{\rm Py}}=0$ (solid black line). This result is obvious as at the limit of well-separated Co and Py stripes, the frequency of modes concentrated in the Py stripes will be the same as in isolated Py stripes. As $l$ decreases, the frequency of the fundamental mode in the Co/Py MC decreases slower than in the Py MC, and at $l$ values between 30 and 250 nm, its  stays even above the fundamental mode of the Py MC for $K_{S,{\rm Py}}\gg\mu_0 M_{S,{\rm Py}}^2 d/4$ (dashed black line -- strong exchange pinning). With the width of the NMS dropping below 30 nm, the Co/Py fundamental mode again moves below the fundamental mode of single-component MC with large $K_{S,{\rm Py}}$, close to the line for Py MC with $K_{S,{\rm Py}}=0$ (solid black line), and finally approaches 0 when $l \rightarrow 0$. The Co stripes are characterized by higher magnetization saturation $M_{\rm S}$ and therefore the SWs occupy them weaker in the bi-component MC for the frequencies lower than the FMR frequency of the isolated Co stripe, given by Kittel formula (\ref{Eq:Kittel}). In fact, the excitation of SWs in Co stripes inside the Co/Py MC is possible only due to the neighbourhood of Py stripes. The penetration of SWs into Co stripes strengths, in turn, the dynamic interaction between the Py stripes.\cite{Huajun18}

In the long-wavelength limit (at $k \approx 0$) when $l=0$, the bi-component MC can be regarded as a homogeneous material with effective properties (i.e. effective magnetization and exchange constant), and thus its fundamental mode frequency can be also described by Eq.~(\ref{Eq:Kittel}).  

 The last observation can be used to explain the experimental results presented in Ref.~[\onlinecite{Wang09}] for a 1D bi-component MC composed of Co and Py stripes. In the experimental report, the frequency of the fundamental mode at the Brillouin zone center is nonzero at $H_0 = 0$. However, with a 2~nm-wide NMS, Fig.~\ref{Fig:bi_comp_gap_dep} shows that the first mode (of frequency $\sim2.5$~GHz) is located at the center of the Brillouin zone. Such a minute separation (of around 1\% of the width of the stripe) may plausibly occur in the sample during its fabrication. Moreover, if the sample is exposed to air, oxidation can introduce  some surface anisotropy at the stripe edge\cite{Gruyters02,Rosa07,Lucari04,Shipton09,Zhu10} resulting in a further shift in the frequency of the fundamental mode due to dynamic magnetization pinning. Hence, the shift in frequency due to the NMSs  and oxidation may explain the disagreement between numerical results and experimental data found in Ref.~[\onlinecite{Sokolovskyy11}]. In actual fact, the 1 nm wide gap was introduced in the micromagnetic simulations in Ref.~[\onlinecite{Zhang12}], where it was admitted, that misalignment during the two-step lithographic process used in the preparation of bi-component samples may result in such a gap at the interface of two alternating magnetic media.

\begin{figure}[ht]
   \includegraphics[width=8cm]{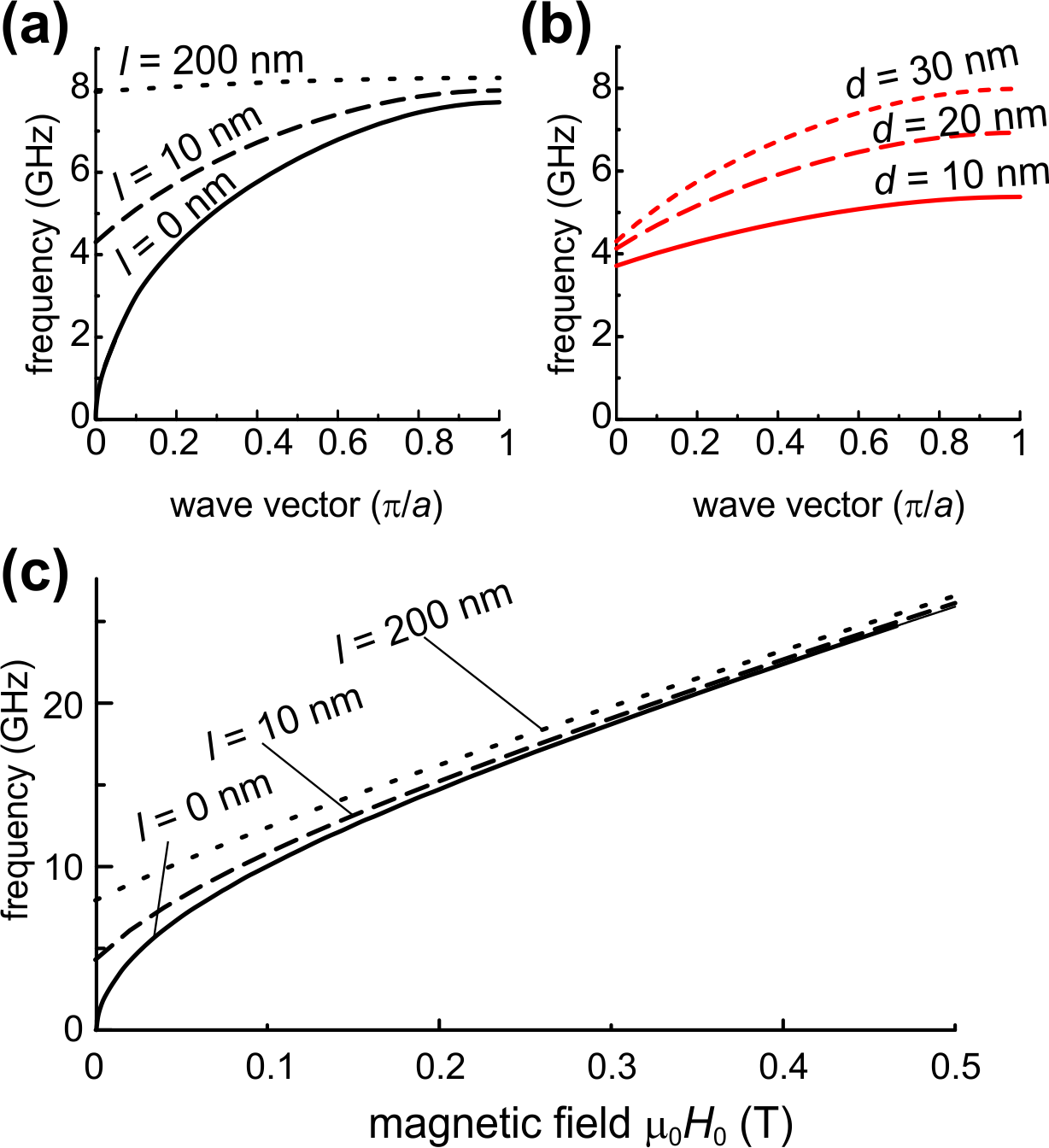}
   \caption{ (a-b) Dispersion relation of SWs in the first Brillouin zone in a bi-component MC with $H_0=0$: (a) for fixed thickness $d = 30$ nm and three separations between the Co and Py stripes -- the solid, dashed and dotted lines correspond to $l=$ 0, 10 and 200 nm, respectively; (b) for a fixed NMSs' width of $l =10$ nm, the solid, dashed and dotted lines correspond to the MC's thicknesses of $d=$  10 nm, 20 nm and 30 nm, respectively. For every separation value in (a), the respective MC has a different lattice constant, and so, the Brillouin zone boundary also appears at a different wave number.
   (c) The fundamental mode frequency at $k=0$ in dependence on the external magnetic field for the three NMSs' widths $l=$ 0, 10 and 200 nm (solid, dotted and dashed lines, respectively).   }
\label{Fig:bi_comp_disp}
\end{figure}

After our discussion of the variation of the FMR frequency with $l$, we also study  the dispersion relation $f(k)$ of SWs in bi-component MCs. To illustrate the relation between the  $f(k)$ dependencies for single- and bi-component MCs, we plotted them together in Fig.~\ref{Fig:bi_comp_gap_dep}(b) for one selected separation between the stripes $l=2$ nm. The dispersion relation for the single-component MC is folded into the first Brillouin zone of the bi-component MC because of our selection of a unit cell composed of two Py stripes. We can notice that the artificial crossing of the dispersion branches for the single-component MC is transformed into an anticossing for the bi-component MC at the edge of the Brillouin zone. Therefore, the bands in the bi-component MC are in general narrower  than those characterizing the single-component MC. We can expect that the width of the first band [$\Delta f = f(k=\pi/a) - f(k=0)$] will decrease with increasing $l$, because for well-separated stripes, the dispersion relation is flat [$f(k=\pi/a) = f(0)$] due to the lack of interaction between the stripes. Indeed, this property is visible in the results of the calculations presented in Fig.~\ref{Fig:bi_comp_disp}(a). It can be seen, that the change of SW frequencies resulting from the separation between stripes [$f(l \rightarrow \infty) - f(l=200 \text{nm})$] decreases as the wave vector increases, from 8.3~GHz for $k = 0$ to 0.57~GHz at the Brillouin zone boundary. This is despite the fact that the increase in $l$ leads to the extension of the lattice constant (we keep the stripes' width fixed) and thereby to the shift of the Brillouin zone boundary to smaller wave numbers $k$. The weak sensitivity of the SW frequency on $l$ at the Brillouin zone boundary can be attributed to the phase variation of the modes. Indeed, for this wave number, the SW phase changes by $\pi$ between the nearest unit cells, and so, the nodes of the SW amplitude are present at the border of the unit cell.

The SW spectrum of the considered MC is plotted as a function of the bi-component MC's thickness ($d$) in Fig.~\ref{Fig:bi_comp_disp}(b). The width of the NMS  is fixed to $l=10$~nm. Increasing the film thickness results in a strong bending of the dispersion curve of the fundamental mode. This result can be explained in terms of the strength of interaction between the stripes. By definition, the exchange interaction is present only in magnetic media and so the only interaction that creates the SW spectrum in the considered case of an array of stripes separated by  NMSs is magnetostatic in nature. Increasing the separation of the magnetic stripes, as well as decreasing the thickness of the film, reduces the magnetostatic interaction between them. Hence, the modes become less dispersive, and the magnonic band becomes flatter, as can be observed in Fig.~\ref{Fig:bi_comp_disp} (a) and (b). In both cases, the change follows the same trend: upon increasing $l$, the frequency at $k = 0$ (bottom of the first magnonic band) is increased, whilst the increasing $d$ shifts the frequency at $k = \pi/a$  (the top of the first magnonic band) to higher values. This shows how the width of the first magnonic band can be effectively changed merely by introducing structural changes in bi-component MCs.

Finally, we study the fundamental mode as a function of the external magnetic field. The results of the calculations are shown in Fig.~\ref{Fig:bi_comp_disp}(c) for the bi-component MC of 30 nm thickness and three NMS widths, which correspond to Co and Py stripes in direct contact (solid black line), separated by 10 nm (dashed-red line) and by 200 nm (dotted-blue line). We see that the influence of $l$ on the fundamental mode decreases with increasing $H_0$ from 7.93 GHz at $H_0 = 0$ to 0.66 GHz at 0.5 T [we are referring here to the difference $f(l=200$~nm) - $f(l=0$)]. This shows that in 1D MCs the contribution of the external magnetic field prevails over that of the dynamic coupling between separated stripes, and that the fundamental mode is most sensitive to the NMS's  width at fields where softening of the modes happens. 

\section{Conclusions}\label{Sec:Conclusions}
In summary, we have numerically studied the spin wave spectra of planar 1D magnonic crystals formed by periodic arrays of one (and two kinds) of thin ferromagnetic stripes separated by a non-magnetic spacers. The influence of the stripe's thickness and non-magnetic material's width on the ferromagnetic resonance frequency has been investigated fully, with the most significant changes found at small separations between the stripes. Interestingly, these changes result not only from the decreasing dipolar interaction between the stripes with increasing separation width, but also are attributed to the lateral pinning of spin waves at the stripe edges. The pinning of the dipolar origin is always present and depends on the shape of the stripe cross-section, but the ferromagnetic resonance frequency is additionally influenced whenever the surface magnetic anisotropy is present at the stripe edges. Than the influence of the volume magnetostatic charges on spin-wave frequency is also revealed. We roughly estimated, that the spin wave pinning at the edges does not play a significant role, as the non-magnetic spacers' width is larger than the stripes' width.

In bi-component magnonic crystals we have shown that a similar dependence on the non-magnetic spacers' width exists. We have also studied the influence of the width of the non-magnetic  spacer on the width of the first magnonic band. We have shown that both increasing the non-magnetic spacers' width and decreasing the thickness of the magnonic crystal decrease the dynamic coupling. 

The results obtained here can help in the interpretation of  experimental results and have significant impact for applications involving magnonic crystals, as they show how the effective response of the single and bi-component magnonic crystals to a homogeneous microwave field can be modified by spin wave pinning. Thus, these properties are  important for the designing of magnetic metamaterials and magnonic devices. 

\begin{acknowledgments}
The research leading to these results has received funding, hereby gratefully acknowledged, from the European Community's Seventh Framework Programme (FP7/2007-2013) under Grant No.~228673 (MAGNONICS), No.~247556 (NoWaPhen), from the European Unions Horizon 2020 research and innovation program under Marie Sklodowska-Curie Grant Agreement No. 644348 (MagIC), and from
the Polish National Science Centre projects No.~UMO-2016/21/B/ST3/00452 and No.~UMO-2017/24/T/ST3/00173.
\end{acknowledgments}

\vspace{1 cm}
\bibliography{bibliography}

\end{document}